\documentclass{pasj00}
\color{black} 

\begin{document}
\SetRunningHead{Y.\ Terada {\it et al.}}{Suzaku Discovery of Hard X-ray Pulsations from AE Aquarii}
\Received{2007/09/18}
\Accepted{2007/11/14}

\title{Suzaku Discovery of Hard X-ray Pulsations from the Rotating Magnetized White Dwarf, AE Aquarii}

\author{Yukikatsu \textsc{Terada},\altaffilmark{1}
Takayuki \textsc{Hayashi},\altaffilmark{2}
Manabu \textsc{Ishida},\altaffilmark{3}
Koji \textsc{Mukai},\altaffilmark{4}}
\author{Tadayasu \textsc{Dotani},\altaffilmark{3}
Shunsaku \textsc{Okada},\altaffilmark{3}
Ryoko \textsc{Nakamura},\altaffilmark{3}}
\author{
Sachindra \textsc{Naik},\altaffilmark{5}
Aya \textsc{Bamba},\altaffilmark{3}
and
Kazuo \textsc{Makishima},\altaffilmark{6,7}}
\altaffiltext{1}{Department of Physics, Science, Saitama University,
Saitama 338-8570}
\altaffiltext{2}{Science of Physics, Tokyo Metroporitan University,\\
1-1 Minami-Osawa, Hachioji-si, Tokyo, 192-0397, Japan}
\altaffiltext{3}{Department of High Energy Astrophysics,\\
Institute of Space and Astronautical Science (ISAS),\\
Japan Aerospace Exploration Agency (JAXA),\\
3-1-1 Yoshinodai, Sagamihara, Kanagawa 229-8510, Japan}
\altaffiltext{4}{Exploration of the Universe Division, Code 660,\\
NASA/GSFC, Greenbelt, MD 20771, USA}
\altaffiltext{5}{Physical Research Laboratory, Ahmedabad 380 009, India}
\altaffiltext{6}{Makishima Cosmic Radiation Laboratory, RIKEN, \\
2-1, Hirosawa, Wako-shi, Saitama 351-0198, Japan}
\altaffiltext{7}{Department of Physics, Graduate School of Science,\\
University of Tokyo, Hongo 7-3-1, Bunkyo-ku, Tokyo 113-0033, Japan}
\email{terada@phy.saitama-u.ac.jp}


\maketitle

\begin{abstract}
The fast rotating magnetized white dwarf, AE Aquarii, was observed with Suzaku,
in October 2005 and October 2006 
with exposures of 53.1 and 42.4 ks, respectively.
In addition to clear spin modulation in the 0.5--10 keV band of 
the XIS data at the barycentric period of $33.0769\pm0.0001$ s,
the 10--30 keV HXD data in the second half of the 2005 observation
also showed statistically significant periodic signals at a consistent period.
On that occasion, the spin-folded HXD light curve exhibited two sharp
spikes separated by $\sim 0.2$ cycles in phase, in contrast to
approximately sinusoidal profiles observed in energies below $\sim$ 4
keV.
The folded 4--10 keV XIS light curves are understood 
as a superposition of those two types of pulse profiles.
The phase averaged 1.5--10~keV spectra can be reproduced by two thermal
components with temperatures of $2.90_{-0.16}^{+0.20}$ keV and
$0.53_{-0.13}^{+0.14}$ keV, but the 12-25 keV HXD data show 
a significant excess above the extrapolated model.
This excess can be explained by either a power-law model 
with photon index of $1.12_{-0.62}^{+0.63}$ or 
a third thermal component with a temperature of $54_{-47}^{+26}$ keV.
At a distance of 102 pc, the 4--30 keV luminosities of 
the thermal and the additional components become $1.7_{-0.6}^{+1.3}$ and 
$5.3_{-0.3}^{+15.3} \times 10^{29}$ erg s$^{-1}$, respectively.
The latter corresponds to 0.09\% of the spin down energy of the object.
Possible emission mechanisms of the hard pulsations are discussed, 
including in particular non-thermal ones.
\end{abstract}

\section{Introduction}
\label{section:introduction}
Since the discovery of cosmic-rays by Hess in 1912, 
the origin of such high-energy particles and 
the mechanism of their acceleration have remained 
a long-standing issue in astrophysics.
Among various types of astrophysical particle acceleration sites,
one important prototype is rotation-powered pulsars,
namely fast rotating neutron stars (NSs)
with strong magnetic field of $\geq 10^{12}$ G.
They are considered to accelerate particles using their rotation 
as the ultimate energy source, and
strong electric fields as the acceleration tool.
The electrostatic potential, induced by the strong surface field, $B$,
and the fast rotation with a period, $P$, is expected to amount to 
\begin{eqnarray}
V &\sim& \left(\frac{2 \pi R}{P}\right) B R \nonumber \\
&\sim& 6 \times 10^{16} \left(\frac{P}{1 \mbox{s}}\right)^{-1}
\left(\frac{B}{10^{12} \mbox{G}}\right)
\left(\frac{R}{10^6 \mbox{cm}}\right)^2 \mbox{V}
\label{eq:acc_volt}
\end{eqnarray}
where $R$ is a typical radius from the NS center
at which the electric acceleration takes place.

Magnetized white dwarfs (WDs) are similar systems to pulsars;
rotating compact objects with strong magnetic fields.
Since a typical magnetized WD has $P \sim 5 \times 10^3$ s, $B \sim 10^6$ G,
and $R \geq 10^9$ cm, 
we expect $V \sim 10^{13}$ V from equation (\ref{eq:acc_volt}).
Therefore, magnetized WDs should be a promising candidate of 
new particle acceleration sites,
although efficiencies of the particle acceleration and 
the subsequent radiation could be different between pulsars and WDs.

Non-thermal incoherent radio emission has in fact been 
detected from seven WDs via systematic radio surveys
\citep{mcv_radio1,mcv_radio2,mcv_radio3,mcv_radio4,mcv_radio5}.
Some objects like AE Aquarii \citep{aeaqr_radio1, aeaqr_radio2} and 
AM Herculis \citep{amher_radio1,amher_radio2} show 
signs of coherent radio emission.
Then, higher-energy photons in the X-ray to gamma-ray bands 
may also be generated via such non-thermal processes as 
curvature radiation like in NSs, non-thermal bremsstrahlung, 
and inverse Compton scattering of soft photons
\citep{nonthermal_process1, nonthermal_process2,nonthermal_process3}, 
as well as electron-positron annihilation lines
and $\gamma$-ray emission via the $\pi^0$ decay chain.
In fact, some authors reported TeV gamma-ray emission 
from AE Aquarii and AM Herculis during optical flares
\citep{aeaqr_tev1, aeaqr_tev2,amher_tev}.
Therefore, high energy electrons with energies reaching 1-100 MeV 
are thought to be present in these WDs \citep{aeaqr_radio2,aeaqr_radio_theo}.

The object AE Aquarii (hereafter AE Aqr), located
at a distance of $d = 102_{-23}^{+42}$ pc \citep{aeaqr_distance},
is a magnetized cataclysmic variable (MCV)
belonging to the DQ Herculis (or intermediate polar) class.
It is spinning near break up \citep{aeaqr_particle_acceleration}
with a spin period of $P_{\rm sp} = 33.0767$ s \citep{aeaqr_optical},
which is the second fastest value among MCVs.
This fast rotation, 
combined with the surface field intensity 
of $B \sim 10^{5 \mbox{--} 6}$ G 
which is a typical value among intermediate polars, 
yields via equation (\ref{eq:acc_volt})
an induced potential of $V\simeq 10^{14 \mbox{--} 15}$ V, 
which is probably one of the highest found among MCVs.
In addition, AE Aqr has been in  a spin-down state over the past 27 years, 
at a rather high rate of $1.8 \times 10^{-6}$ s yr$^{-1}$ 
\citep{aeaqr_spindown,aeaqr_spindown2}. 
The value implies a huge spin-down luminosity of
$6 \times 10^{33}$ erg s$^{-1}$ 
\citep{aeaqr_particle_acceleration,aeaqr_spindown}.
The object is one of the most asynchronous cataclysmic variables
with an orbital period of $P_{\rm orb} = 35567~{\rm s} = 9.88$ hr 
\citep{wd_period,aeaqr_spindown}.
Flares reported in the radio band \citep{aeaqr_radio1,aeaqr_radio2} 
are consistent with synchrotron emission \citep{aeaqr_radio_theo},
and the pulsating TeV gamma-rays \citep{aeaqr_tev1,aeaqr_tev2} are 
sometimes coincident with optical flares.

According to soft X-ray observations \citep{aeaqr_asca1, aeaqr_newton},
the plasma temperature of thin thermal radiation from AE Aqr
is significantly lower than the estimated gravitational potential of its WD,
whose mass is about 0.8 -- 0.9 times the solar mass 
\citep{wd_period,aeaqr_orbital_spin_param}.
Furthermore, its UV to X-ray luminosity ($10^{31}$ erg s$^{-1}$) is
three orders of magnitude lower than that in a simple accretion case
\citep{mcv_mdot} with a rate of $10^{17}$ g  s$^{-1}$.
Therefore, the accreting matter may be inhibited 
from reaching the WD surface by the fast WD rotation \citep{aeaqr_newton}. 
\citet{aeaqr_propeller} argue that only 1 \% of the accreting gas
in AE Aqr reaches the WD surface, 
while the rest is expelled away due to propeller effects.
These observed properties altogether suggest strongly that
AE Aqr is a promising site of particle acceleration,
as pointed out by \citet{aeaqr_particle_acceleration}.

The hard X-ray band is most suited to the search 
for the expected non-thermal signals,
because that particular band is expected to receive the largest number of
non-thermal high-energy photons without being hampered 
by the soft X-rays arising from accretion.
In the present paper, 
we report on the first successful detection of pulsating hard X-rays 
from the  most promising object, AE Aqr, 
achieved with Suzaku \citep{suzaku07},
which carries two instruments; the X-ray CCD camera called 
the X-ray Imaging Spectrometer (XIS;\cite{xis2007}) 
with the X-ray telescope (XRT; \cite{xrt2007}) and 
the Hard X-ray Detector (HXD;\cite{hxd2007a}).
The emission is likely to be of non-thermal origin,
thus reinforcing the view that MCVs are indeed particle accelerators. 

\section{Observation and Data Reduction}
\label{section:observation}
\subsection{Suzaku Observation of AE Aqr}
\label{section:observation:suzaku}
We observed AE Aqr with Suzaku twice 
during the performance verification phase.
The first observation was performed from 2005 October 30 02:13 UT
to November 2 01:02 UT (sequence number 400001010),
and the second one from 2006 October 25 5:25 UT 
to 09:11 UT on the next day  (sequence number 400001020).
The observations were carried out at the ``HXD nominal'' pointing position 
(i.e., with the target on the nominal optical axis of the HXD).
Although the HXD has no imaging capability,
the full field of view of its PIN diodes (1.0 degrees square)
contained no hard X-ray objects in the Swift BAT catalog.
On both occasions, the XIS was operated in the normal mode 
with the ``1/8 window'' option,
which gives a time resolution of 1 s, without charge injection function.
The HXD was operated in the nominal mode;
in the second observation, however, 
a quarter of  the 64 PIN diodes were operated 
with a reduced voltage of 400 V, rather than the nominal 500 V.

\subsection{Data Reduction}
\label{section:observation:reduction}
We used the datasets produced by the Suzaku pipe-line 
processing version 1.2.2.3 with the calibration-version of
hxd20060829, xis20060913, xrt20060720, and xrs20060410,
using tools in HEADAS package version 6.2.
We used XSPEC version 11.3.2, to fit spectra with models.

The source was detected with the XIS in the 0.5 -- 10 keV band
at an intensity of 0.37 c s$^{-1}$ and 0.28 c s$^{-1}$ per sensor, 
in 2005 and 2006, respectively. 
In the XIS analysis, we excluded all telemetry-saturated data portions,
and data taken in ``low'' data-rate mode.
We further removed those intervals
when the source elevation above Earth's limb was below $5^{\circ}$,
Earth's day-night boundary was less than $25^{\circ}$,
or the spacecraft was in, and 436 s after, leaving 
the South Atlantic Anomaly (SAA).
We then accumulated nominal-grade events within 6 mm ($4\farcm3$)
of the image centroid. 
The XIS background events were taken from 
a source-free region of the corresponding sensor.
The derived background,
exhibiting a rate of $\sim 0.02$ c s$^{-1}$ per sensor in 0.5--10 keV, 
is generally negligible except in the highest end of the XIS bandpass.

In calculating XIS ancillary response files, 
we took into account efficiency reduction caused 
by the window option and 
by un-controlled fluctuations of the optical axis of the XRT \citep{xrt2007}.
In the first observation, 
the effective areas  of XIS-0, 1, 2, and 3 were
thus reduced to  92.3\%, 93.1\%, 93.4\%, and  93.7\%
of their nominal values, respectively,
while those during the second observation were  
93.7\%, 93.2\%, 92.8\%, and 93.8\% respectively.

The HXD data were screened using the same criteria as used on the XIS data.
In addition, we discarded data 
acquired during time intervals when the geomagnetic
cutoff rigidity was lower than 8 GV.
After these filtering steps, 
the final event list of HXD-PIN was obtained using only those events 
that survived the standard anti-coincidence function of the HXD (DET\_TYPE=1). 
The non X-ray background (NXB) of HXD-PIN was synthesized 
by appropriately combining night-Earth data sets
acquired under different conditions \citep{hxd2007b}, 
using METHOD=`PINUDLC' and version of METHODV=1.2 \citep{HXD_NXB_reports}.

After selecting event lists, the first observation yielded 
net exposures of about 70.3 and 53.1 ks 
with the XIS and the HXD, respectively.
Those in the second observation were 
48.6 with the XIS, and 42.4 ks with the HXD.

\section{Timing Analyses}
\label{section:timing}
\subsection{Light Curves}
\label{section:timing:lc}
\begin{figure}[htb]
\centerline{
\FigureFile(0.95\columnwidth,){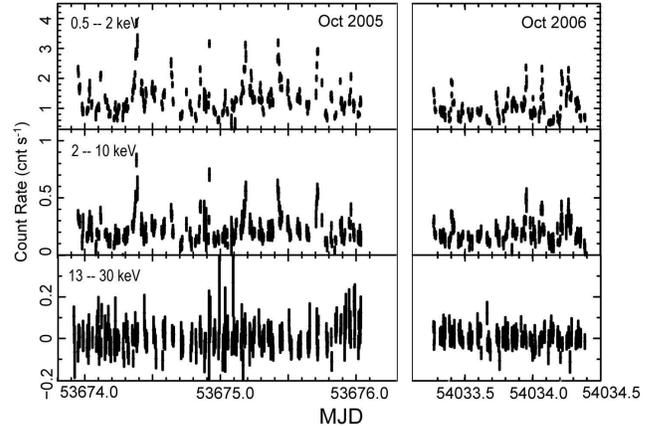}
}
\caption{Background-subtracted light curves of AE Aqr during the two observations, obtained with the XIS (0.5 -- 2 keV and 2 -- 10 keV) and the HXD (13 -- 30 keV). The HXD data are corrected for dead time. The XIS and HXD-PIN data are binned into 250 s.}
\label{fig:lc}
\end{figure}

Background-subtracted light curves of AE Aqr, 
obtained with the XIS and the HXD, 
are shown in figure~\ref{fig:lc}. 
From this figure, it is not obvious 
whether the source is detected with HXD-PIN.
More quantitatively, the background-subtracted 13--30 keV PIN counts 
integrated over 1 ks become
$14.0 \pm 14.5 \pm 4.0$ c ks$^{-1}$ and $ 18.0 \pm 20.9 \pm 6.0$ c ks$^{-1}$, 
in 2005 and 2006, respectively;
where the first uncertainties are statistical 90\% errors,
while the second ones are systematic errors in the PIN NXB subtraction
estimated from the reproducibility of counts 
during the satellite pointing at Earth.
For reference, the PIN NXB counts in the same energy range 
were 198 and 202 c ks$^{-1}$ in 2005 and 2006, respectively.
Therefore, the source detection with HXD-PIN becomes 
significant both in the 2005 and 2006 observations
when integrated over the full exposure of $\sim 50$ ks.
However, the detection becomes marginal 
when we consider contribution from the cosmic X-ray background (CXB)
which is estimated to be 18.0 c ks$^{-1}$ in the 13--30 keV PIN band,
based on its canonical surface brightness \citep{Boldt1987} and 
the angular response of the PIN diodes to uniform illumination.
Aiming at a higher detection sensitivity, 
below we search the data for periodic modulations
synchronized with the WD rotation,
before carrying out spectral analyses.

In both observations, the XIS data reveal several flares
involving  two- to three-fold intensity increases,
each lasting for $\sim 2500$ s.
In the PIN data, however, 
no corresponding  flux increases are found,
primarily because of insufficient count statistics.
Even if PIN counts increased by a factor of 3 in 2500 s 
(maximum $\sim 38$ c per 25 ks),
the associated statistical errors, $\sim 36$ c, would be comparable.

\subsection{XIS Period Determination}
\label{section:timing:XISperiod}
After applying the standard barycentric corrections
to the photon arrival times  \citep{hxd_timing}, 
we searched the XIS light curves extracted from the source region
for periodic intensity variations synchronized with the WD rotation.
In doing so, we did not subtract the background,
because the background events occupy
only 5.0\% of the total count rate of the XIS, and 
no periodicity around $P_{\rm sp}$ was found in the background data.
The standard epoch-folding analysis of the 0.5--10 keV XIS data
revealed highly significant source pulsations on both occasions,
at a barycentric period of $P_{\rm XIS}=33.0769$ s.
After \citet{period_err}, 
68\% errors associated with these period determinations are estimated to be 
0.0063 s and 0.0104 s, in 2005 and 2006, respectively.
Within these errors, the derived period is consistent
with the values of $P_{\rm sp}$ measured in the optical 
\citep{aeaqr_optical,aeaqr_spindown}, 
in X-rays with Ginga, ASCA \citep{aeaqr_asca_ginga}, 
XMM-Newton \citep{aeaqr_newton}, and Chandra \citep{aeaqr_spindown2},
and in TeV $\gamma$-rays \citep{aeaqr_tev1}.
Therefore, we identity $P_{\rm XIS}$ 
with the rotational period of the WD in AE Aqr;
that is, $P_{\rm XIS} = P_{\rm sp}$.

\begin{figure}[htb]
\centerline{
\FigureFile(0.95\columnwidth,){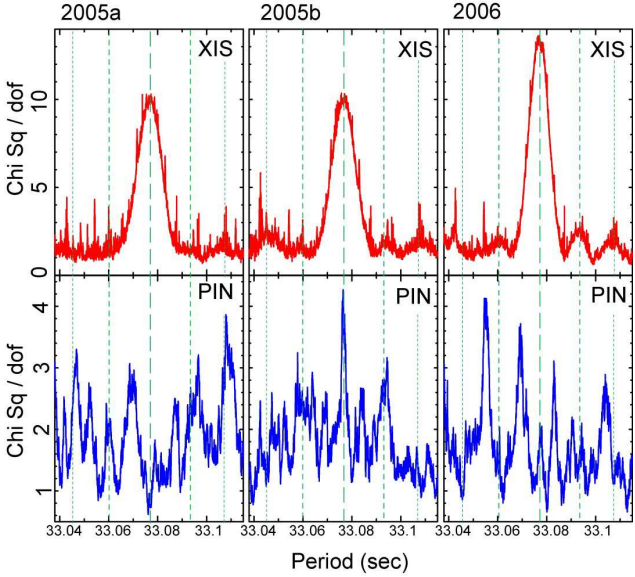}
}
\caption{Periodograms calculated from the background-inclusive 
XIS data (red) and the HXD-PIN data (blue).
The vertical axis shows reduced $\chi^2$, 
calculated against a hypothesis that the folded data profile is constant.
The period step is $2.5 \times 10^{-5}$ s,
using 7 and 64 bins per cycle for the XIS and HXD, respectively.
The vertical green lines indicate various expected periods;
the spin period $P_{\rm SP}$ (thick dashed);
$P_{\rm pro}$ and $P_{\rm ret}$ (thin dashed);
and the beat periods between $P_{\rm SP}$ and 1 day (dotted).
(left) Results from the first half of the 2005 data,
where the PIN energy band is 10 -- 30 keV.
(middle) The same as left panel, 
but for the second half of the 2005 observation.
(right) Results from the entire 2006 observation,
where the PIN energy band refers to 13 -- 30 keV.}
\label{fig:periodogram}
\end{figure}

Figure~\ref{fig:periodogram} (top) shows 0.5--10 keV XIS periodograms,
used in determining $P_{\rm XIS}$.
Strictly speaking, the above period determinations utilized
the full length of the 2005 and 2006 data sets,
while the periodograms in this figure were obtained 
with the 2005 data divided into the first and second halves 
(hereafter 2005a and 2005b, respectively),
each having a 90.2 ks gross duration,
and a net exposure of  26.4 and 34.6 ks, respectively.
The very similar shapes of the three (2005a, 2005b, and 2006) XIS periodograms
indicate that the periodic modulation in the XIS band is quite stable.
Although we could in principle determine the period with a much higher 
accuracy by coherently combining the 2005 and 2006 datasets,
the periods on individual occasions are too uncertain to extrapolate 
them for one year without pulse-count ambiguities.
Instead, we incoherently combined the 2005 and 2006 periodograms, 
which reduced the period error to 0.0001s \citep{period_err}. 
We hence quote the XIS-determined period as 
$P_{\rm XIS} = 33.0769 \pm 0.0001$ s.

In energies below 2 keV, 
the pulse-peak phase comes at BJD (Barycentric Julian Ephemeris Day)
2453673.5000134 $\pm$ 0.0000011,
from the combined dataset of 2005 and 2006 
folded on the period of $P_{\rm XIS}$.
When this epoch and the above determined $P_{\rm XIS}$ are employed, 
our measurement predicts the pulse peak to occur at BJD 2453613.2767025 
during the Chandra observation conducted on 2005 August 31.
This agrees, within 0.0000033 days, or 0.29 s with that actually 
measured on that occasion by \citet{aeaqr_spindown2}, 
namely BDJ 2453613.2767058 $\pm$ 0.0000039.

\begin{figure}[hbt]
\centerline{
\FigureFile(0.95\columnwidth,){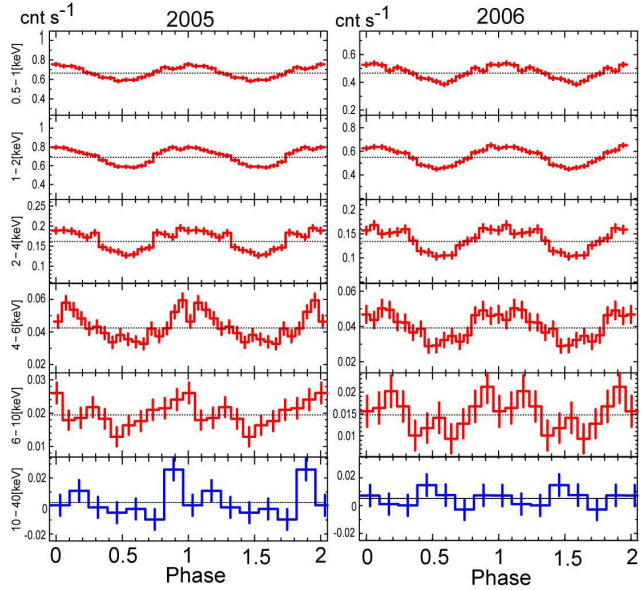}
}
\caption{Energy-resolved light curves folded 
on the XIS-determined period of 33.0769 s.
The non X-ray backgrounds are subtracted from the XIS data, 
but not from the HXD-PIN data. 
The energies from top to bottom panels are 
0.5 -- 1 keV, 1 -- 2 keV, 2 -- 4 keV, 4 -- 6 keV, 
and 6 -- 10 keV for the XIS, while 10 -- 40 keV or 13 -- 40 keV 
with the HXD for the 2005 and 2006 observations, respectively. 
The vertical axis shows the count rate in each energy band, 
with the maximum scale set to 1.3 times the averaged count rate of that band.
Phase 0.0 corresponds to BJD 2453673.5000, and 
the date of the pulse peak in the 0.5-2 keV band
(at a spin phase of 0.035) occurs on
BJD 2453673.5000134 $\pm$ 0.0000011.}
\label{fig:eresolved_efold}
\end{figure}

The XIS light curves, folded on the spin period $P_{\rm XIS}$,
are presented in figure~\ref{fig:eresolved_efold} in several energy bands.
Thus, the modulation at lower energies has a sinusoidal profile,
with an amplitude of $\sim \pm 16\%$ of the mean intensity.
Toward higher energies, the folded pulse profiles become more structured;
this may be responsible for the fine structures 
seen in the XIS periodograms (figure~\ref{fig:periodogram}).
Further analysis of the folded pulse profiles is 
conducted in subsection \ref{section:timing:pulse}.

\subsection{HXD Period Search}
\label{section:timing:HXDperiod}
After the same barycentric corrections,
we folded background-inclusive PIN data on the period of $P_{\rm XIS}$.
The 2005 data were analyzed in an energy range of 10--30 keV,
while those of 2006 in 13--30 keV, because thermal noise arose
around the PIN lower threshold level in this observation.
Since the PIN data have much poorer signal statistics,
we tentatively divided the period into seven phase bins.
Then, as shown in figure~\ref{fig:eresolved_efold} (bottom) in blue,
a statistically significant modulation with a spiky profile
was found from the 2005 data
with a reduced chi-squared of $\chi^2_\nu=2.5$ for $\nu=6$, whereas
the 2006 data was consistent with being constant ($\chi^2_\nu=0.62$).

Now that the PIN signals exhibit a hint of periodic modulation 
at the period of $P_{\rm XIS}$ at least in the first observation,
we proceed to more systematic epoch-folding analyses on 
the background-inclusive 10--30 keV (or 13--30 keV) PIN light curves.
The reason for not subtracting the background is
that the actual background would not vary significantly 
in 30 s \citep{hxd2007b},
and that the synthetic PIN background models have
larger uncertainties when used on time scales of $\sim 1$ s,
as would be needed here.
Specifically, we calculated HXD periodograms
over a trial period range of 33.04 to 33.12 s,
with a typical period step of $2.5\times 10^{-5}$ s or even finer.
The bin number per cycle was varied in the range between 7 and 13.
Furthermore, to avoid false quantization effects
due to finite phase bin numbers, 
we calculated, at each trial period, chi-squares 
by changing the time origin within a single phase bin,
and then selected the maximum value to construct the periodogram.
Typical results from these studies
are shown in figure~\ref{fig:periodogram} in blue,
in comparison with those from the XIS data.
Thus, the three (2005a, 2005b, and 2006) PIN periodograms
all exhibit multiple peaks reaching $\chi^2_\nu= 3-4$.
Therefore, the PIN signals are inferred to be significantly 
variable on times scales of $\sim 33$ s,
possibly involving rather rapid variations
as suggested by the narrow widths of these peaks.

The most important finding from these studies is
that the 2005b PIN periodogram exhibits its highest peak
with $\chi^2_\nu=4.2$ (for $\nu=6$)
at a period of $P_{\rm HXD} =33.0764 \pm 0.0005$ s,
which agrees well with $P_{\rm sp}$, 
and hence $P_{\rm XIS}$ within respective errors.
The probability of this peak arising by chance is lower than $7\times 10^{-7}$.
Although the results presented here were obtained
using a particular 7 phase bins per cycle,
the results do not change 
if different numbers of bins (e.g., 8, 11, or 13) are utilized. 
The reason why the PIN profile in figure~\ref{fig:eresolved_efold}
exhibited $\chi^2_\nu= 2.5$ instead of the peak value ($\chi^2_\nu= 4.2$)
is that it was calculated at $P_{\rm XIS}$ rather than $P_{\rm HXD}$.
The peak at  $P_{\rm HXD}$ does not disappear 
even if we divide the 2005b data into two halves
each comprising 32 PIN detectors.
Furthermore, side lobes seen at 33.060 s and 33.092 s in the 2005b plot
can be interpreted as due to beat between $P_{\rm sp}$ and 
the Suzaku sampling window.
These facts suggest that 
the periodic signal at $P_{\rm HXD}$ in the 2005b PIN data is real,
and is intrinsic to the WD rotation.
The narrower peak in the PIN periodogram compared to that of the XIS, 
is consistent with the folded PIN profile having sharp structures,
as already suggested by figure~\ref{fig:eresolved_efold}.
In contrast, the PIN  signal modulation at $P_{\rm sp}$ 
is not significant either in the 2005a or the 2006 data.

\begin{figure}[hbt]
\centerline{
\FigureFile(0.85\columnwidth,){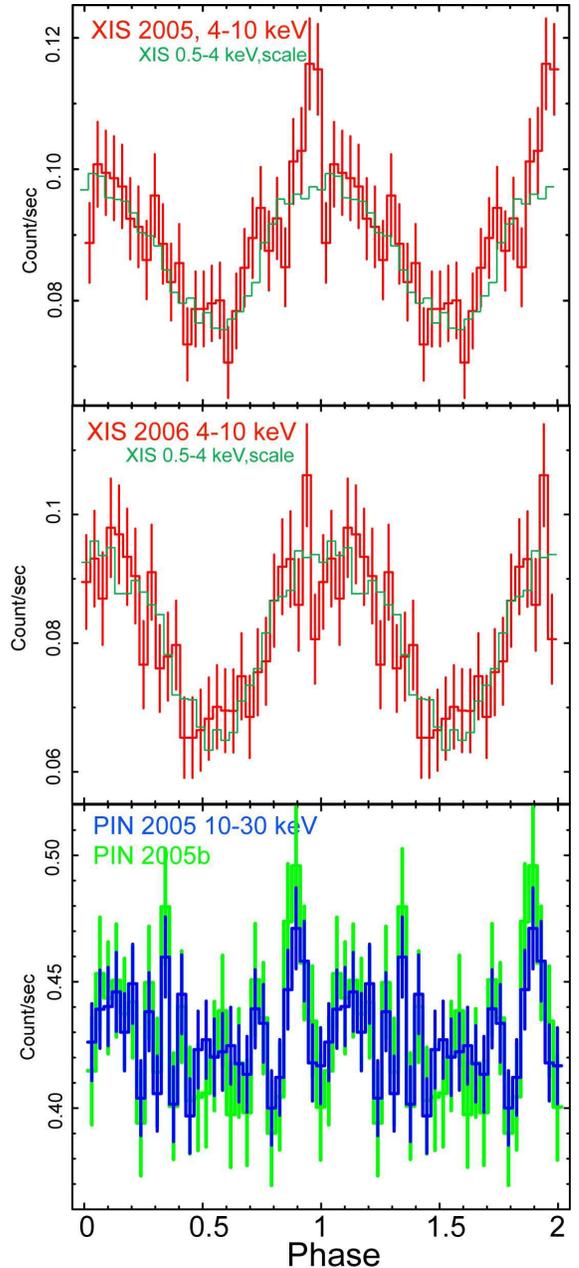}
}
\caption{
Soft and hard X-ray profiles of AE Aqr,
folded on the HXD-determined period $P_{\rm HXD}$.
Phase 0.0 corresponds to BJD 2453673.5000.
See the text for the relative XIS vs. HXD timing alignment.
(Top and middle) Background-inclusive 0.5--4 keV (green) 
and 4 -- 10 keV  (in red) XIS profiles acquired in 2005 and 2006, respectively.
The 0.4--4 keV profile is arbitrarily scaled to approximately agrees 
with the 4 -- 10 keV profile.
(Bottom) The background-inclusive 10--30 keV HXD-PIN data,
taken in the entire 2005 observation 
in blue and the second half of 2005 in green.}
\label{fig:efold_dbl_pulse}
\end{figure}

\subsection{Pulse Profiles}
\label{section:timing:pulse}
We have so far obtained promising evidence
that the PIN signals are indeed modulated at $P_{\rm sp}$.
Of course, the peak in the 2005b periodogram (figure~\ref{fig:periodogram})
could be an accidental coincidence 
caused by some non-periodic hard X-ray variations,
considering the presence of multiple peaks in the other PIN periodograms.
However, in figure~\ref{fig:eresolved_efold}, the XIS profiles become
similarly structured toward higher energies.
Therefore, we expect that the hard X-ray periodicity can be reinforced
by comparing the folded PIN profiles with those from the XIS 
in harder energies.

We sorted the XIS data into two bands (0.5--4 keV and 4--10 keV),
and folded them both at the period of $P_{\rm HXD}$.
For reference, we also folded the 2005 HXD-PIN data in 10--30 keV 
at the same period, with the same bin size as for the XIS data,
namely 29 bins per cycle.
The folded profiles derived in the three energy bands 
are presented in figure~\ref{fig:efold_dbl_pulse}.
As already suggested by figure~\ref{fig:eresolved_efold},
the PIN profiles indeed exhibit several sharp spikes, 
among which at least two, 
one at a pulse phase of $\phi \sim0.9$ and the other at $\phi \sim0.1$,
are statistically significant.
As shown in figure~\ref{fig:efold_dbl_pulse} bottom,
the 2005 PIN signal modulation becomes stronger 
when we limit the data to the second half (2005b) 
in agreement with the inference from 
the periodogram (figure~\ref{fig:periodogram});
the modulation amplitude of the spike changes from 
0.03 $\pm$ 0.01 c s$^{-1}$ to 0.05 $\pm$ 0.01 c s$^{-1}$.
Importantly, a similar spike is also observed at $\phi \sim 0.95$ 
in the 4--10 keV XIS profile in 2005 (figure~\ref{fig:efold_dbl_pulse}),
superposed on the approximately sinusoidal intensity modulation
which dominates the spin-folded profiles in lower energies.
As seen in figure~\ref{fig:efold_dbl_pulse} (middle),
the sharp features in the 4 -- 10 keV XIS profile is found also from
the 2006 data, but with a reduced amplitude, in agreement with 
the negative detection of the periodic signal in the 2006 PIN data 
(section \ref{section:timing:HXDperiod}).

\begin{figure}[hbt]
\centerline{
\FigureFile(0.8\columnwidth,){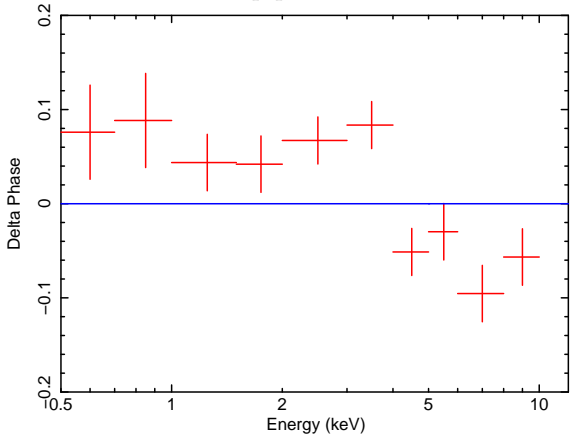}
}
\caption{Energy dependence of the phase of the XIS pulse profiles from the 2005 observation, relative to the 10 -- 30 keV PIN profile. The vertical axis shows the phase at which the cross correlation value of each profile to the PIN profile becomes maximum. The folded XIS profiles refer to figure~\ref{fig:eresolved_efold} (left), while that of HXD-PIN the 2005 data in figure~\ref{fig:efold_dbl_pulse} (bottom). }
\label{fig:e_dependence_efold}
\end{figure}

In figure~\ref{fig:efold_dbl_pulse}, 
the major sharp spike shows 
a slight phase difference by $\phi \sim 0.05$ 
between the XIS (4--10 keV) and PIN data sets.
This could be either real or instrumental, 
because the relative timing between the XIS and HXD 
is accurate only to within $\sim \pm 0.5$ s
even though the absolute timing of the HXD has been verified 
to an accuracy of 360 $\mu$s \citep{hxd_timing}.
In order to study at which energy this spike appears, 
and to examine whether the major spike exhibits 
any energy-dependent phase shift,
we cross-correlated the energy-sorted XIS profiles of the 2005 observation
(figure~\ref{fig:eresolved_efold} left) against 
the pulse template from the PIN data 
(i.e., figure~\ref{fig:efold_dbl_pulse} bottom).
The results, shown in figure~\ref{fig:e_dependence_efold}, 
reveals a phase discontinuity at about 4 keV.
This means that the XIS profile below 4 keV is 
dominated by the sinusoidal modulation,
and the separate spike starts appearing above 4 keV.
Since the cross-correlation peak is thus found consistently at 
$\phi \sim -0.06$ over the 4 -- 10 keV energy band of the XIS,
the apparent phase difference (by $\phi \sim 0.05$) seen in 
figure~\ref{fig:efold_dbl_pulse} between the XIS (4--10 keV) and PIN profiles
is likely to be instrumental rather than real.

Based on the close similarity between the 4--10 keV XIS folded profile
and those from HXD-PIN, we conclude that the periodicity in the PIN signal 
(at least in 2005b) is in fact due to the WD rotation,
and hence AE Aqr emits, at least occasionally, 
and at least in the 4-30 keV band,
a periodic hard X-ray emission component whose folded profile 
consists of a few sharp pulses.

\subsection{Other issues}
\label{section:timing:others}
While we have so far concentrated on the signal modulation
at the spin period $P_{\rm sp}$,
the signals could also be modulated at 
beat periods between the rotational and orbital periods,
namely at so-called prograde period 
$P_{\rm pro}=(P_{\rm sp}^{-1} - P_{\rm orb}^{-1})^{-1}$ =  33.1075 s,
or retrograde period
$P_{\rm ret}=(P_{\rm sp}^{-1} + P_{\rm orb}^{-1})^{-1}$ =  33.0460 s.
If, for instance, the WD is rotating in the same sense as its binary motion,
and the X-ray emission region is stationary when observed 
from a frame of reference co-rotating with the binary,
the signals would be modulated at $P_{\rm pro}$ rather than at $P_{\rm sp}$.
Indeed, the 2005a PIN periodogram (figure~\ref{fig:periodogram} left)
exhibits the highest peak  at $33.1080\pm 0.0003$ s,
which is close to $P_{\rm pro}$.
All the XIS periodograms (figure~\ref{fig:periodogram}) also
show some enhancements at  $\sim P_{\rm pro}$.
However, the peak in the 2005a PIN periodogram
is slightly (but significantly) offset from the exact $P_{\rm pro}$.
Moreover, the 2005b PIN data exhibit
no significant modulation  at $P_{\rm pro}$, nor at $P_{\rm ret}$.
Therefore, we 
conclude that the hard X-ray periodicity at the beat periods are 
less convincing, and defer further examination to a future publication.

\begin{figure*}[hbt]
\centerline{
\FigureFile(1.45\columnwidth,){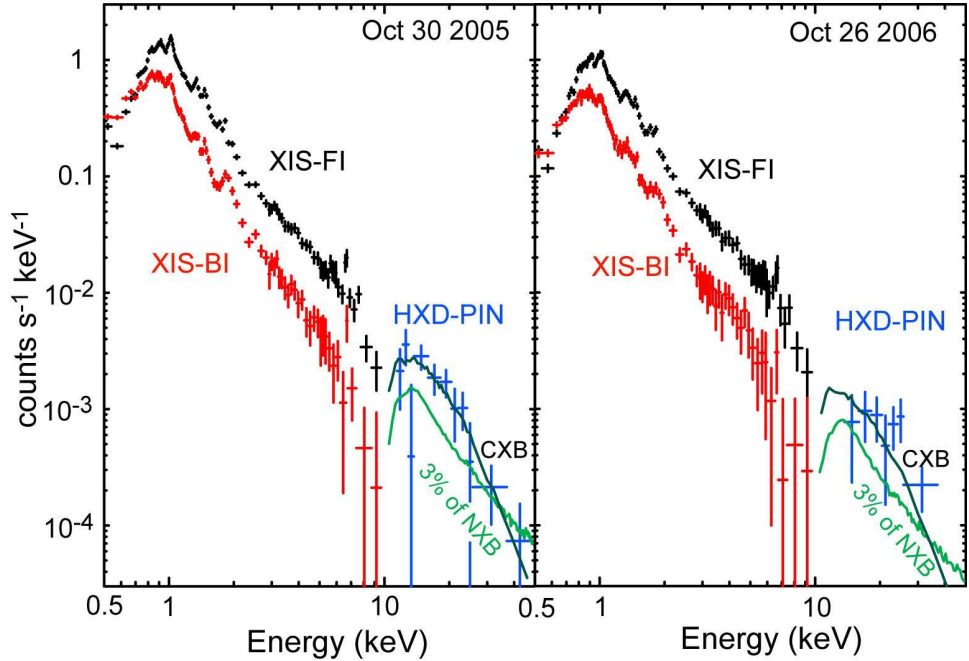}
}
\caption{Background-subtracted spectra of AE Aqr obtained with Suzaku in 2005 and 2006 (left and right, respectively) 
presented without removing the instrumental responses; error bars are statistical only. The light green curves show the typical uncertainty of the PIN non X-ray background, and the dark green curves indicates the expected cosmic X-ray background \citep{Boldt1987}.}
\label{fig:average_spec}
\end{figure*}

Since the WD in AE Aqr has a projected orbital semi-major axis 
of $\sim$ 2 l-s around the binary center of gravity,
the X-ray pulse arrival times are expected to vary
by this amount depending on the orbital phase.
In fact, \citet{aeaqr_spindown2} successfully detected this effect, 
and determined the orbital modulation amplitude as $A=2.17 \pm 0.48$ l-s;
this is $\pm 8\%$ of the pulsation period.
While this may be negligible when the pulse profile is sinusoidal,
the orbital effects could have significant effects
when it has sharp structures, as is in fact the case with the 2005b PIN data.
Therefore, we attempted to correct the arrival times, $t$, 
of individual X-ray events by subtracting 
the expected orbital modulation,
\begin{equation}
\Delta t = A \cos \left[ 2\pi \left( \frac{t-t_0}{P_{\mbox{orb}}} - \xi \right) \right],
\label{equation:orbital_correction}
\end{equation}
where $t_0$ is the time of the superior conjunction of the WD,
and $\xi~(0 \le \xi \le 1)$ is any possible phase shift.

Employing the nominal orbital parameters by \citet{aeaqr_spindown}, 
namely $\xi = 0.0$, $t_0$ = BJD 2445172.2784, and $A = 2.04$ s,
where $A$ is consistent within errors ($2.04 \pm 0.13$ s)
with the value $2.17 \pm 0.48$ s by \citet{aeaqr_spindown2},
we applied the correction of equation~(\ref{equation:orbital_correction})
to the 4--10 keV XIS data.
Then, the periodogram peak at $P_{\rm XIS}$ became indeed higher,
with $\chi^2/\nu$ increasing from 7.8 to 9.2 in 2005, 
and from 6.5 to 8.0 in 2006.
When the same correction is applied to the  PIN data,
the 2005b periodogram peak at $P_{\rm HXD}$ has also increased from 4.2 to 4.3,
although no major effects were observed in the other two PIN periodograms.
This reinforces our inference 
that the hard X-ray pulsation  in the 2005b data is due to the WD rotation.
When we  allowed $A$ and $\xi$ to vary,
the 4--10 keV XIS data preferred $A=1.5 \pm 0.5$ and 
$\xi = -0.1 \pm 0.1$,
which are not much different from the nominal ephemeris.
Although errors are large,
the 2005b PIN data also favor these parameter regions.

\section{Spectral Analyses}
\label{section:spectrum}
To study the origin of the pulsating hard X-ray component revealed 
in the timing analyses (section~\ref{section:timing}),
we analyzed the pulse phase-averaged spectra taken with the XIS and the HXD.
In this analyses, we discarded half the PIN data acquired in 2006, 
under the reduced voltage of 400V,
because the PIN response under this non-standard operation
condition is not yet fully calibrated.

\begin{figure}[h]
\centerline{
\FigureFile(0.7\columnwidth,){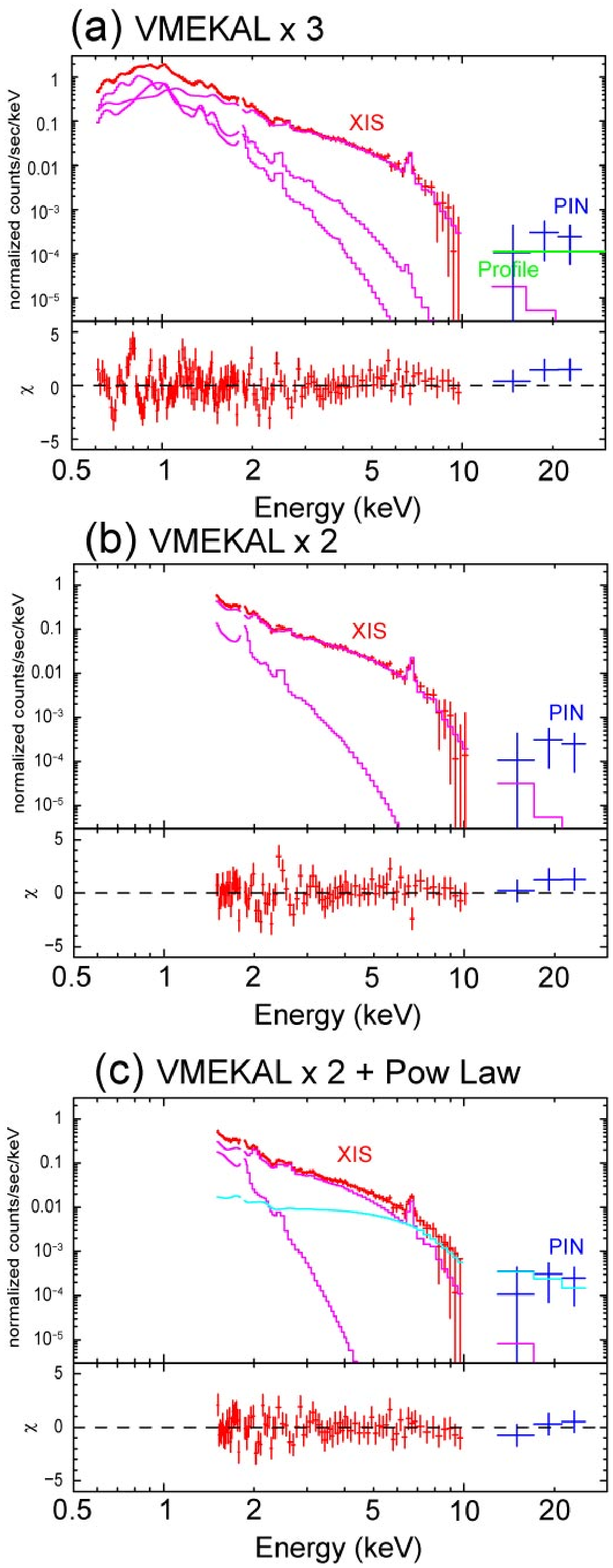}
}
\caption{Model fits to the phase-average Suzaku spectra, 
summed over the two observations (2005 and 2006). 
Red and blue crosses show the XIS and PIN data, respectively, 
from which the NXB and CXB have been subtracted. 
(a) A fit with a three-temperature VMEKAL model in the 0.5 -- 25 keV band, 
with the parameters listed in table~\ref{table:average_spec}
(labeled as 2005+2006). 
The horizontal green line indicates the 13--30 keV PIN X-ray count rate
obtained from the folded pulse profile 
(figure~\ref{fig:efold_dbl_pulse} bottom),
as a difference between $\phi=$ -0.1 -- 0.1 and $\phi=$ 0.1 -- 0.9.
(b) A fit with a two-temperature VMEKAL model, in the 1.5 -- 25 keV band. 
The best-fit parameters are listed in table~\ref{table:average_spec2}. 
(c) The same as panel (b), but a power-law with photon index of 1.12 (cyan) 
is added.}
\label{fig:average_spec_fit}
\end{figure}

The background-subtracted X-ray spectra of AE Aqr 
obtained in this way are shown in figure~\ref{fig:average_spec}.
In agreement with the inference made in section~\ref{section:timing:lc},
the signal detection with the HXD is significant both in 2005 and 2006,
up to $\sim$ 25 keV. 
This statement remains valid even if we consider the systematic errors 
in the PIN NXB subtraction, which is estimated to be $\sim 3$ \%
in the present case as judged from the reproducibility 
of the Earth occultation data.
However, as shown in figure~\ref{fig:average_spec},
the PIN detection becomes marginal when we consider the CXB contribution.
Therefore, we further subtracted the CXB from the spectra, 
and summed the result over two observations.
As plotted in figure~\ref{fig:average_spec_fit}a,
we obtained positive PIN signals 
at a level of $\sim 2\times 10^{-4}$ c s$^{-1}$ kev$^{-1}$ around 20 keV.
Thus the source has been detected up to $\sim 25$ keV.
There, the green line in the figure represents 
the phase-averaged 13 -- 30 keV count rate 
of the spiky signals, $\sim 1.5\times 10^{-4}$ c s$^{-1}$ keV$^{-1}$,
obtained as a difference between the pulse-peak phase ($\phi$ = -0.1 -- 0.1)
and the other phase in figure~\ref{fig:efold_dbl_pulse} (bottom). 
The good agreement in figure~\ref{fig:average_spec_fit}a,
between the spectral and timing results from HXD-PIN,
indicates that the emission in the PIN range is pulsating nearly 100\%.

The phase averaged X-ray spectra of AE Aqr has been shown 
with ASCA and XMM-Newton \citep{aeaqr_asca_ginga,aeaqr_newton}
to be reproduced by an optically thin thermal plasma emission model
with a few different temperature components.
Thus, to reproduce the Suzaku spectra in the 0.5 -- 25 keV band,
we adopted a multi-temperature VMEKAL model \citep{mekal,mekal2}
which are constrained to have common metal abundances
but allowed to have separate temperatures.
These components were subjected to a common photoelectric absorption
with a free neutral column density.
The overall model normalization was constrained to be the same
between the XIS and PIN.
In the fitting, we ignored the systematic errors in the PIN NXB estimation,
because the PIN flux was consistent with the pulse flux in the spin profile
as already described.
As summarized in table~\ref{table:average_spec} (noted as ``2005+2006''),
the spectra were roughly reproduced by three VMEKAL components,
and the derived abundances are consistent with 
those by XMM-Newton \citep{aeaqr_newton}.
However, the fit was unacceptable with 
large reduced chi-squared values of 1.79,
and was not improved even if we added another VMEKAL component.
In addition, the model fell an order of magnitude short of the PIN data.

\begin{table*}
\caption{Best fit parameters of the multi temperature VMEKAL Model to the phase averaged spectra.}
\label{table:average_spec}
\begin{center}
\begin{tabular}{lcccc}
\hline 
{}   &{XMM-Newton 2001$^*$}    &{2005}    &{2006}   &{2005+2006}\\
\hline\hline 
{$N_{\mbox{H}}$ ($10^{20}$cm$^{-2}$)}   
                  &{$3.59_{-1.20}^{+1.47}$}&{$<1.28$}               &{$<2.04$}               &{$<1.92$}\\
{$kT_1$ (keV) }   &{$4.60_{-0.47}^{+0.60}$}&{$3.17_{-0.15}^{+0.25}$}&{$3.19_{-0.35}^{+0.65}$}&{$3.04_{-0.22}^{+0.14}$}\\
{$kT_2$ (keV) }   &{$1.21_{-0.08}^{+0.13}$}&{$0.93_{-0.08}^{+0.05}$}&{$0.87_{-0.13}^{+0.07}$}&{$0.86\pm0.06$}\\
{$kT_3$ (keV) }   &{$0.59\pm0.02$}         &{$0.53_{-0.03}^{+0.02}$}&{$0.50_{-0.07}^{+0.04}$}&{$0.50_{-0.05}^{+0.04}$}\\
{$kT_4$ (keV) }   &{$0.14_{-0.02}^{+0.05}$}&{--}&{--}&{--}\\
\hline 
{N (Abundance$^\dagger$)}    &{$3.51_{-0.81}^{+0.92}$}&{3.51 fix}&{3.51 fix}&{3.51 fix}\\
{O (Abundance$^\dagger$)}    &{$0.74_{-0.23}^{+0.17}$}&{$0.66_{-0.10}^{+0.09}$}&{$0.57\pm0.16$}&{$0.64_{-0.05}^{+0.09}$}\\
{Ne (Abundance$^\dagger$)}   &{$0.43_{-0.25}^{+0.28}$}&{$0.66\pm0.17$}         &{$0.49_{-0.18}^{+0.26}$}&{$0.68_{-0.16}^{+0.14}$}\\
{Mg (Abundance$^\dagger$)}   &{$0.70_{-0.14}^{+0.15}$}&{$0.76_{-0.13}^{+0.10}$}&{$0.68_{-0.18}^{+0.22}$}&{$0.77\pm0.12$}\\
{Si (Abundance$^\dagger$)}   &{$0.81_{-0.12}^{+0.14}$}&{$0.65_{-0.11}^{+0.09}$}&{$0.43_{-0.12}^{+0.13}$}&{$0.58_{-0.10}^{+0.09}$}\\
{S (Abundance$^\dagger$)}    &{$0.73_{-0.18}^{+0.20}$}&{$0.89\pm0.15$}         &{$0.70\pm0.19$}         &{$0.80\pm0.15$}\\
{Ar (Abundance$^\dagger$)}   &{$<0.89$}               &{$0.67_{-0.43}^{+0.45}$}&{$0.31_{-0.41}^{+0.66}$}&{$0.55_{-0.46}^{+0.47}$}\\
{Ca (Abundance$^\dagger$)}   &{$<1.11$}               &{$0.37_{-0.37}^{+0.53}$}&{$0.83\pm 0.80$}        &{$0.68_{-0.60}^{+0.59}$}\\
{Fe (Abundance$^\dagger$)}   &{$0.47_{-0.06}^{+0.07}$}&{$0.40_{-0.05}^{+0.04}$}&{$0.33\pm 0.07$}        &{$0.39_{-0.02}^{+0.04}$}\\
{Ni (Abundance$^\dagger$)}   &{$1.27_{-0.50}^{+0.57}$}&{$1.36_{-0.39}^{+0.37}$}&{$0.89_{-0.42}^{+0.54}$}&{$1.14_{-0.34}^{+0.31}$}\\
\hline 
{Flux$^\ddagger$($10^{-12}$ erg s$^{-1}$ cm$^{-2}$)} &{11.6}&{$6.83_{-0.26}^{+0.33}$}&{$6.22_{-0.24}^{+0.38}$}&{$6.65_{-0.29}^{+0.20}$}\\
\hline 
{$\chi^2_\nu$(dof)} &{1.22(992)}&{1.94(152)}&{1.45(152)}&{1.79(152)}\\
\hline 

\multicolumn{5}{@{}l@{}}{\hbox to 0pt{\parbox{85mm}{\footnotesize
       \par\noindent
       \footnotemark[$*$] From \citet{aeaqr_newton}.
       \par\noindent
       \footnotemark[$\dagger$] Solar abundances by \citet{abundance_anders}.
       \par\noindent
       \footnotemark[$\ddagger$] In the 0.5 -- 10 keV band.
}\hss}}

\end{tabular}
\end{center}
\end{table*}

The large value of reduced chi-squared from the three-temperature VMEKAL fit
(table~\ref{table:average_spec}) is
mainly caused by disagreements between the model and the data 
around 1 keV, where the Fe-L line complex is present.
Thus, we ignored the data below 1.5 keV and fitted the 1.5 -- 25 keV data 
with two VMEKAL components,
whose abundances were fixed to the values by \citet{aeaqr_newton}
for simplicity.
As shown in figure~\ref{fig:average_spec_fit}b and
listed in table~\ref{table:average_spec2}, 
the fit became much improved with reasonable temperatures of 
$2.90_{-0.16}^{+0.20}$ keV and $0.53_{-0.13}^{+0.14}$ keV.
However, the model still disagrees with the PIN data points.

To reproduce the PIN flux, we added an additional power-law component 
to the two VMEKAL model, and obtained an acceptable result 
as shown in figure~\ref{fig:average_spec_fit}c
and table~\ref{table:average_spec2}.
The F-statistic of this improvement is 15.5,
which means that the probability of this improvement being caused
by chance is $2.0\times 10^{-6}$. 
Thus, adding another component is justified.
The power-law modeling is, however, not unique, and
another VMEKAL model instead of the power-law model also provides a
significantly improved fit at a temperature of $54_{-47}^{+26}$~keV
(table~\ref{table:average_spec2}), with the F-test value of 11.1 or 
the chance probability of $3.5\times 10^{-6}$.
The derived temperature is essentially considered a lower limit,
i.e., $> 7$ keV. Since a thermal component of such a temperature is
not rare among accreting WDs, we cannot distinguish, 
from the spectral analysis alone, 
whether the additional hard X-ray component detected with the HXD is 
thermal or non-thermal.

\begin{table*}
\caption{Best Fit Parameters of the multi temperature VMEKAL$^\dagger$ with/without Power Law model to the phase averaged spectra in the 1.5 -- 25 keV band.}
\label{table:average_spec2}
\begin{center}
\tabcolsep 3pt
\begin{tabular}{ccccccccc}
\hline 
{Model$^*$}&
{$N_{\mbox{H}}$$^\dagger$}&{$kT_1$$^\ddagger$}&{Flux$^\S$}
                &{$kT_2$$^\ddagger$}&{Flux$^\S$}
                &{Photon Index or $kT_3$$^\ddagger$} & {Flux$^\S$} 
                &{$\chi^2_\nu$ (dof)}\\
\hline 
{i}&
{$<0.165$}    &{$2.90_{-0.16}^{+0.20}$}&{$3.77_{-0.18}^{+0.30}$} 
              &{$0.53_{-0.13}^{+0.14}$}&{$3.18_{-0.97}^{+4.61}$} 
              &{--}&{--}
              &{$1.27$ (84)}\\
{ii}&
{$0.66_{-0.38}^{+0.36}$}
              &{$2.04_{-0.21}^{+0.09}$}&{$1.96_{-0.37}^{+0.32}$} 
              &{$0.27_{-0.05}^{+0.05}$}&{$0.74_{-0.56}^{+1.64}$} 
              &{$1.12_{-0.62}^{+0.63}$}&{$0.59_{-0.05}^{+1.80}$}
              &{$0.95$ (82)}\\
{iii}&
{$0.67_{-0.20}^{+0.17}$}
              &{$2.01_{-0.23}^{+0.17}$}&{$2.13_{-0.64}^{+0.03}$} 
              &{$0.26_{-0.04}^{+0.07}$}&{$3.82_{-2.83}^{+6.86}$} 
              &{$54.4_{-47.2}^{+25.7}$}&{$0.66_{-0.20}^{+11.9}$}
              &{$0.95$ (82)}\\
\hline 

\multicolumn{5}{@{}l@{}}{\hbox to 0pt{\parbox{155mm}{\footnotesize
       \par\noindent
       \footnotemark[$*$] Abundances are fixed to the XMM-Newton values \citep{aeaqr_newton}.
Model (i) is double VMEKAL, (ii) is double VMEKAL with power law, 
and (iii) is triple VMEKAL mode. 
       \par\noindent
       \footnotemark[$\dagger$] Column density in $10^{20}$ cm$^{-2}$.
       \par\noindent
       \footnotemark[$\ddagger$] Temperature in keV.
       \par\noindent
       \footnotemark[$\S$] Flux in $10^{-12}$ erg s$^{-1}$ cm$^{-2}$ in 0.5 -- 10 keV.
}\hss}}

\end{tabular}
\end{center}
\end{table*}

\section{Discussion}
\label{section:discussion}
\subsection{Summary of the Observation}
\label{section:discussion:obs}
Using Suzaku, we observed AE Aqr in 2005 and 2006.
In the first observation, the HXD detected periodic signals synchronized
with the rotation of the WD (section \ref{section:timing:HXDperiod})
whose profile has a sharp peak (section \ref{section:timing:pulse}) with
a duration of $\sim$0.1 in phase.
The same sharp feature was also found in the
spin-folded XIS light curves in energies above 4 keV.
The phase-averaged spectra from the XIS above 1.5~keV can be explained
with the two optically thin thermal plasma models with temperatures of
$2.90_{-0.16}^{+0.20}$~keV and $0.53_{-0.13}^{+0.14}$~keV
(section \ref{section:spectrum}).
The PIN spectrum up to 25~keV, however, shows 
an excess emission over the model.
This excess can be fitted by a power-law model 
with a photon index of $1.1 \pm 0.6$, or another thermal model
with the temperature of $54_{-47}^{+26}$ keV.
In the 13--30 keV range, the spectrally measured excess,
$3.2^{+9.7}_{-2.5}$ c s$^{-1}$, is consistent within large errors
with that contained in the sharp peaks 
in the spin-folded hard X-ray light curves.

The 4--30 keV flux of the additional hard component
from the spectral analyses is obtained as
$L_{\rm HX} \sim 1.7_{-0.1}^{+4.9}\times 10^{-12}$ erg cm$^{-2}$ s$^{-1}$,
which gives a luminosity of $5.3_{-0.3}^{+15.3} \times 10^{29}$ erg s$^{-1}$
at the distance of 102~pc \citep{aeaqr_distance}.
This corresponds to $\sim 0.01$\% of the spin-down luminosity of
$6 \times 10^{33}$ erg s$^{-1}$
\citep{aeaqr_particle_acceleration, aeaqr_spindown}.
Note, however, that the hard X-ray pulsation has 
a duty ratio of only $\sim$0.1. 
If the hard X-ray emission is highly anisotropic without a constant bias, 
and if the emission is beamed away from us in the other pulse phases, 
the true hard X-ray luminosity would be higher, 
amounting up to $5 \times 10^{30}$ erg s$^{-1}$
which is $0.1$\% of the spin-down luminosity. 
In any case, 
the ratio of the hard X-ray emission to the spin-down luminosity 
of AE Aqr comes in the range of 0.01--0.1\%, which is the same as 
those measured in the 2--10 keV range from NS pulsars 
\citep{ns_spindown1, ns_spindown2}.

\subsection{The Nature of the Hard X-ray Component}
\label{section:discussion:hard_component}
The observed hard X-ray emission pulsates at the WD rotation period,
but not at the prograde or retrograde beat periods.
In order to reinforce this result, we further folded the 4--10 keV 
XIS data at the prograde and retrograde beat periods, 
but the results were consistent with the absence of any periodic modulation,
either in the 2005 or 2006 data sets.
This implies that the hard X-ray emission region
is fixed in the frame which is co-rotating with the WD,
while excludes a possibility that the emission region
is fixed to the binary frame.

Although approximately in phase with the soft X-ray spin modulation, the
hard X-ray pulses exhibit a spiky profile with a duty ratio of only
$\sim 0.1$ in phase, unlike the sinusoidal profile in the XIS band below
4 keV.
Furthermore, the spin phase of the major spike exhibits an energy
dependence with a discontinuity at $\sim$ 4 keV
(figure~\ref{fig:e_dependence_efold}).
The soft and hard X-ray components are therefore likely 
to originate from distinct emission sites. 
If interpreted as hot thermal emission,
the hard component, with its very high temperature,
should be emitted from the bottom of the gravitational potential,
i.e., the WD surface, 
while the softer component from some outer regions \citep{aeaqr_newton}.
However, it would be highly difficult to produce the observed sharp pulse 
profiles from the WD surface (e.g., magnetic poles) via any thermal process.
Indeed, such X-ray pulsations, appearing only in the hard X-ray band 
with very sharp profiles, have never been observed 
from other accreting magnetic WDs \citep{mcv_asca2004}, 
in which the thermal emission mechanism is considered to dominate.
We thus consider that the hard component is more naturally interpreted 
as non-thermal emission, as evidenced by its successful power-law modeling.
If so, a population of non-thermal particles must be accelerated 
in the AE Aqr system, at least occasionally, producing 
the hard X-ray pulses detected with Suzaku.
In the following two subsections,
we discuss possible particle acceleration mechanisms,
and hard X-ray emission mechanisms from the accelerated electrons.

\subsection{Possible Acceleration Mechanism}
\label{section:discussion:acc_mechanism}
As mentioned in section \ref{section:introduction}, 
the fast rotating WD in the AE Aqr system is expected to produce 
an induced electric potential reaching $V\simeq 10^{14 \mbox{--} 15}$ V.
Therefore, it is most natural to assume that 
the non-thermal particles are acceleration by this electric potential.

Although the suggested acceleration mechanism is 
very similar to that in rotation-powered pulsars, 
the present WD system differs from pulsars in an important physical condition.
While a rotation-powered pulsar is expected 
to reside in a low-density environment, 
AE Aqr is an accreting binary system.
In particular, the radius of the light cylinder,
$l_{\rm lc} \sim P_{\rm sp} /2\pi \sim 5$~l-s, 
is comparable to the binary size of $\sim 6$ l-s.
In fact, \citet{aeaqr_newton}, 
analyzing the XMM-Newton RGS data of AE Aqr,
concluded that a plasma with the electron density 
$n_{\rm e} \sim 10^{11}$~cm$^{-3}$ 
is spread over a scale of $(2-3)\times 10^{10}$~cm in this binary system.
Since this plasma density exceeds the Godreich-Julian density 
\citep{goldreich69} which is calculated as $\sim 5\times10^{4}$ cm$^{-3}$ 
in the present case \citep{nonthermal_process3},
the induced potential would be short-circuited, and hence 
an efficient acceleration would not take place.
However, both theoretical \citep{aeaqr_propeller} and observational works 
suggest that the propeller effect produces a low-density cavity around the WD,
by inhibiting the accretion plasma from getting closer to the WD 
than $\sim 10^{10}$ cm \citep{aeaqr_propeller}.
The electric acceleration may be possible in such regions.

Even if the plasma density is sufficiently low, 
an efficient electric-field acceleration also requires 
evacuation of neutral matter from the acceleration region;
otherwise, the electrons would suffer too heavy ionization losses
to be accelerated efficiently.
The mean-free-path of an electron with energy $E_{\rm e}$,
due to ionization loss, is given by
\begin{eqnarray}
\lambda_{\rm e} &\sim& \frac{1}{n_{\rm p}} \frac{1}{\ln \Lambda}
\left( \frac{\pi e^2}{8 \epsilon_0 E_{\rm e}} \right)^{-2} \nonumber \\
&\sim& 10^{7} \left( \frac{n_{\rm p}}{10^{10} \mbox{cm$^{-3}$}}\right)^{-1} 
\left( \frac{E_{\rm e}}{ 1 \mbox{keV}}\right)^2 \mbox{cm},
\end{eqnarray}
where $n_{\rm p}$ is the density of protons,
$\ln \Lambda$ is a Coulomb logarithm (typically $\sim$ 10 -- 20),
$e$ is the elementary charge, and 
$\epsilon_0$ is the vacuum dielectric constant 
(see e.g., \cite{rybichi_lightman}).
If an initial electron energy is $E_{\rm e} \sim$ 1 keV 
assuming a thermal plasma in the system,
the energy gain of the electron accelerated 
along the length of $\lambda_{\rm e}$ 
can be roughly estimated to be 
$E_{\rm acc} \sim \frac{V}{l_{\rm lc}} \cdot \lambda_{\rm e} 
\sim 10^{10} \left( \frac{n_{\rm p}}{10^{10} \mbox{cm$^{-3}$}}\right)^{-1}$ eV,
with $V \sim 10^{14}$ V and $l_{\rm lc} \sim 10^{11}$ cm.
Thus, the energy gain $E_{\rm acc}$ exceeds the initial energy $E_{\rm e}$
when the surrounding density is $n_{\rm p} < 10^{17}$ cm$^{-3}$.
Therefore, 
in the AE Aqr system with $\sim 10^{11}$ cm$^{-3}$ \citep{aeaqr_newton}, 
the ionization loss does not affect the electron acceleration.

\subsection{The Hard X-ray Emission Mechanism}
\label{section:discussion:emission_mechanism}
In general, there are the following four elementary processes that can
produce hard X-ray emission from accelerated electrons: 
non-thermal bremsstrahlung, inverse Compton scattering, 
synchrotron emission, and curvature radiation. 
Considering the observational results, we examine these processes. 

If the hard X-rays are due to non-thermal bremsstrahlung in thick
materials, electrons have to be accelerated only to 
an energy of $\sim$ 100 keV to emit X-rays with an energy of 
several tens of keV.
The emanating hard X-ray spectrum is expected 
to have a photon index of $\sim$1, when the target is thick and hence 
the electrons have a flattened energy distribution 
due to Coulomb losses \citep{loss_flattened_bremss}.
This is in good agreement with the observed photon index,
$1.12^{+0.63}_{-0.62}$.
Candidates of the dense matter, co-rotating with the WD,
includes the surface of the WD and 
the accretion stream toward the magnetic poles.
However, to produce the sharp hard X-ray pulses,
the electrons must be in the form of collimated beams
(e.g., along the magnetic field lines), and furthermore, 
must be moderately relativistic ($\geq$ 200 keV)
so that the bremsstrahlung emissivity is sufficiently anisotropic
\citep{bremsstrahlung} to produce the spike in the hard X-ray light curve.
It is not obvious whether these conditions can be fulfilled.
Furthermore, a still larger difficulty is in energetics:
in the 10 -- 100 keV range,
an electron loses $\sim 10^{4}$ times larger energies in 
ionization (or Coulomb) loss than in radiation.
Then, the particle acceleration requires an energy input
which is more than an order of magnitude higher than
the spin-down luminosity. We therefore conclude 
that the bremsstrahlung interpretation is difficult.

In the inverse-Compton scenario,
the seed photons can be either the UV photons from the WD or the optical
photons from the companion star whose spectral type is K3--K5 
\citep{aeaqr_orbital_spin_param}.
Then, the electrons are required to have 
a Lorentz factor of $\gamma \sim$10$^2$, or an energy of $\sim 100$ MeV.
The estimated 4--30 keV luminosity (after accounting for the duty ratio;
section \ref{section:discussion:obs}) translates to 
a hard-photon number flux of $2.6 \times 10^{37}$ s$^{-1}$.
This is a reasonably small fraction ($\sim 10^{-7}$) of the optical photon
number flux to be expected from a K4 star 
($\simeq 5 \times 10^{44}$ s$^{-1}$ near the WD surface). 
However, the inverse-Compton process is rather inefficient;
an electron with a Lorentz factor $\gamma \simeq 10^{2}$ is expected
to lose half its energy in $T_{\rm 1/2} \simeq 
\frac{1}{\gamma}\frac{m_{\rm e}c^2}{\sigma_{\rm T} c u_{\rm s}} 
\sim 4 \times 10^{6}$ s, where $m_{\rm e}$ is the electron mass, 
$\sigma_{\rm T}$ is the Thomson cross section,
and $u_{\rm s} \simeq 0.1$ erg cm$^{-3}$ is the visible photon density
at the position of the WD.
Therefore, the electron is expected to leave the system
after depositing only $\sim 10^{-6}$ of its energy.
The energy input needed to sustain the particle acceleration would be
then $\sim 10^{6} L_{\rm HX}$, 
making the inverse Compton scenario highly unlikely.

As the third possibility, we consider synchrotron radiation
by relativistic electrons.
For a magnetic field strength of $B = B_5 \times 10^{5}$ G,
an electron is required to have a Lorentz factor 
$\gamma \sim 10^4 B_5^{-1}$ to emit $\sim$ 30 keV photons.
This value of $\gamma$ is lower than the maximum possible energy 
expected from the induced voltage (section~\ref{section:introduction}) 
by $\sim$4 orders of magnitude. 
Furthermore, the synchrotron emission is highly efficient,
with a very short lifetime of 
$\sim 8 \times 10^{-6} B_5^{-2} (\frac{\gamma}{10^4})^{-1}$ s 
(e.g, \cite{synchrotron,synchrotron2}).
Considering further that the synchrotron emission is highly anisotropic,
we presume that this scenario is quite likely,
if the electrons are accelerated along the field lines, 
and then at some stage, acquire cross-field velocity
and start emitting synchrotron photons.
The observed photon index $\Gamma \sim 1.1$ predicts a very flat electron 
spectrum with a spectral index of $\sim$1.2. 
This does not matter, however, if direct electron field acceleration works.
In fact, rotation-powered NS pulsars generally exhibit rather
flat spectra in X-rays, $\Gamma \sim 1.4$ \citep{Pulsar_index01}.

Finally, we consider the possibility 
of the curvature radiation process. 
According to \citet{curvature_radiation},
the energy of the hard X-ray emission via curvature radiation can be written as
\begin{equation}
E \sim \frac{3}{2} \frac{\hbar c \gamma^3}{R_{\rm crv}}
\sim 50 \left(\frac{\gamma}{10^6}\right)^3 
\left(\frac{R_{\rm crv}}{R_{\rm WD}}\right)^{-1} \rm{keV},
\label{eq:curvature}
\end{equation}
where $\hbar$ is the Plank constant,
$R_{\rm crv}$ is the curvature radius of magnetosphere field lines,
and $R_{\rm WD}$ is the WD radius of AE Aqr.
On the right-hand side of equation~\ref{eq:curvature}, 
$R_{\rm crv}$ is normalized for convenience to $R_{\rm WD}$.
It is not easy to predict where the photon energy $E$ is maximized,  
because a smaller curvature radius can be realized closer to the WD, 
whereas the maximum $\gamma$ is probably achieved at a larger radius.
Generally, as we move away from the WD surface,
$R_{\rm crv}$ decreases gradually,
while $\gamma$ of the electrons being accelerated will increase rapidly.
According to \citet{nonthermal_process3},
the curvature-radiation photons will reach the maximum of 
$E \sim 100$ keV at $R_{\rm crv} \sim l_{\rm lc}$,
together with $\gamma \sim 6 \times 10^6$,
which is feasible in the AE Aqr system.
However, the efficiency of this process is so low
that the luminosity is estimated to be $\sim 3 \times 10^{27}$ erg s$^{-1}
\sim 10^{-2} L_{\rm HX}$.
Therefore, it is difficult to explain the observation 
in terms of the curvature radiation.

We conclude that the hard X-ray pulse is probably caused by relativistic
electrons accelerated by the fast rotation of the magnetized WD. 
The most promising emission mechanism is the synchrotron emission
in the strong magnetic fields of the WD.

\section*{Acknowledgements}
The authors would like to thank all the members of
the Suzaku Science Working Group,
for their contributions in the instrument preparation,
spacecraft operation, software development,
and in-orbit instrumental calibration.
They are also grateful to the referee, Dr.\ Christopher Mauche,
for very helpful and constructive comments.

\bibliographystyle{pasj}
\bibliography{aeaqr_suzaku_pasj.bib}

\end{document}